\documentstyle[aps,psfig]{revtex}
\begin{document}
\newcommand{\Fstar}{\!\stackrel{*}{F}{\!\!}}
\newtheorem{lemma}{Lemma}
\renewcommand{\thelemma}{\Alph{lemma}}
\newcommand{\beq}{\begin{equation}}
\newcommand{\eeq}{\end{equation}}
\preprint{ICI-EFEI-002/99}
\draft
\title{Singularities in General Relativity coupled to nonlinear electrodynamics}
\author{M. Novello%
\thanks{Electronic mail: \tt novello@lafex.cbpf.br},  
S. E. Perez Bergliaffa, and
J. M. Salim}
\address{Centro Brasileiro de Pesquisas F\'{\i}sicas \\
Rua Dr.\ Xavier Sigaud 150, Urca 22290-180 Rio de Janeiro, RJ -- Brazil} 
\date{\today}
\renewcommand{\thefootnote}{\fnsymbol{footnote}}
\twocolumn[
\hsize\textwidth\columnwidth\hsize\csname@twocolumnfalse\endcsname 
\maketitle

\begin{abstract}
\hfill{\small\bf Abstract}\hfill\smallskip
\par
We study here some consequences of the nonlinearities of the electromagnetic field acting as a source of Einstein's equations on the propagation of photons. We restrict to the particular case of a ``regular black hole'', and show that there exist singularities in the effective geometry. These singularities may be hidden behind a horizon or naked, according to the value of a parameter. Some unusual properties of this solution are also analyzed.  
\end{abstract}
\smallskip\mbox{}]
\footnotetext[1]{Electronic mail: \tt novello@lafex.cbpf.br}
\renewcommand{\thefootnote}{\arabic{footnote}}

\section{Introduction}

It is a well-known fact that some of the most important solutions of Einstein's field equations ({\em e.g.} Friedmann-Robertson-Walker and Schwarzschild)  are singular. However, our understanding of the nature of these singularities is still incomplete. For instance, the cosmic censorship conjecture was put forward by R. Penrose in 1969 \cite{penrose}, but there is still no general proof of it. As a consequence of this lack of understanding, solutions 
that are everywhere regular
and share some of the properties of singular solutions deserve attention.  
This is precisely the case of the ``regular black hole'' spacetimes recently exhibited in \cite{prl,eloy2,eloy3}. 
These solutions were obtained for a very special type of source: an electric field that obeys a 
nonlinear electrodynamics. The authors of \cite{prl} analyzed some of the features of the solution, but left aside others that are relevant. We shall re-examine this solution in detail. More importantly,  
we shall show in this particular example the far-reaching consequences 
of the fact that in nonlinear electromagnetism 
photons {\em do not} propagate along null geodesics of the background geometry. 
They propagate instead along null geodesics of an effective geometry, which depends on
the nonlinearities of the theory. This result, derived by Pleba\'nsky for Born-Infeld electrodynamics \cite{pleb}, was generalized for any nonlinear theory by Guti\'errez {\em et al} \cite{pleb3}, and later independently rediscovered by Novello {\em et al} \cite{nov1}.
Let us mention that 
the propagation of photons beyond Maxwell electrodynamics 
has been studied in several different situations. It has been investigated 
in curved spacetime, as a consequence of non-minimal coupling 
of electrodynamics with gravity \cite{nov2,Drumond,Novello}, 
and in nontrivial QED vacua as an effective modification 
induced by quantum fluctuations \cite{Latorre,Dittrich,Shore}. 
Nearly always, these analysis have had some unexpected results.  
As an example, let us mention the possibility 
of {\em faster and slower-than-light} photons \cite{Latorre}.

Our main concern in this article will be then to show that one must consider the modifications on the trajectories of the photons induced by the nonlinearities of the electromagnetic theory in order to give a complete characterization of spacetimes with a nonlinear electromagnetic source. 
The structure of the paper is the following.  A summary of the the solution given in \cite{prl} and the properties studied there will be given in Section \ref{solution}, along with some interesting properties that went unnoticed before. In Section \ref{geometry} we briefly review 
the origin of the effective geometry for photons in nonlinear electrodynamics. We shall use in Section \ref{anal}
the method of the effective geometry to study the features of the structure that photons see  when travelling in the geometry given in \cite{prl}.
We close with some conclusions.

\section{Details of the solution}
\label{solution}
Ay\'on Beato and Garc\'{\i}a \cite{prl} have found an exact solution of Einstein's equations in the presence of a nonlinear electromagnetic source. The relevant equations  are derived from the action
\cite{pleb2}
\begin{equation}
{\cal S}=\int d^4 x\left[ \frac 1{16\pi }R-\frac 1{4\pi }{\cal L}(F)\right] ,
\label{action}
\end{equation}
where $R$ is the curvature scalar and ${\cal L}$ is a nonlinear function of $F\equiv 
\frac 14F_{\mu \nu }F^{\mu \nu }$. Following \cite{pleb} and \cite{prl}
this system could also be described using another function obtained by means of a Legendre
transformation: 
\begin{equation}
{\cal H}\equiv 2F{\cal L}_F-{\cal L}.  \label{eq:Leg}
\end{equation}
(${\cal L}_F$ denotes the derivative of ${\cal L}$ w.r.t. $F$).
With the definition 
\beq
P_{\mu \nu }\equiv {\cal L}_FF_{\mu \nu },
\label{iden}
\eeq
it can be shown that ${\cal H}$ is a function of $P\equiv \frac 14P_{\mu \nu }P^{\mu \nu }=({\cal L%
}_F)^2F$, {\rm i.e.}, $d{\cal H}=({\cal L}_F)^{-1}d(({\cal L}_F)^2F)={\cal %
H}_PdP$. With the help of ${\cal H}$ one could express the nonlinear
electromagnetic Lagrangian in the action (\ref{action}) as ${\cal L}=2P%
{\cal H}_P-{\cal H}$, which depends on the anti--symmetric tensor $P_{\mu \nu }$%
. The solution of Einstein's equations coupled to nonlinear electrodynamics obtained in \cite{prl} was derived from the following 
source:
\begin{equation}
{\cal H}(P)=P\,\frac{\left( 1-3\sqrt{-2\,q^2P}\right) }{\left( 1+\sqrt{%
-2\,q^2P}\right) ^3}-\frac 3{2\,q^2s}\left( \frac{\sqrt{-2\,q^2P}}{1+\sqrt{%
-2\,q^2P}}\right) ^{5/2},  \label{eq:H}
\end{equation}
where $s=|q|/2m$ and the invariant $P$ is a negative quantity. The
corresponding Lagrangian is given by 
\begin{eqnarray}
{\cal L} & = & P\,\frac{\left( 1-8\sqrt{-2\,q^2P}-6\,q^2P\right) }{\left( 1+\sqrt{%
-2\,q^2P}\right) ^4}- \nonumber\\
& & \frac 3{4\,q^2s}\frac{(-2\,q^2P)^{5/4}\left( 3-2\sqrt{%
-2\,q^2P}\right) }{\left( 1+\sqrt{-2\,q^2P}\right) ^{7/2}}.  \label{eq:Lagex}
\end{eqnarray}
From Eq.(\ref{action}) we get the following equations of motion:
\begin{equation}
G_\mu ^{~\nu }=2({\cal H}_PP_{\mu \lambda }P^{\nu \lambda }-\delta _\mu
^{~\nu }(2P{\cal H}_P-{\cal H})),  \label{Ein}
\end{equation}
\begin{equation}
\nabla _\mu P^{\alpha \mu }=0.  \label{max}
\end{equation}
This system was solved in \cite{prl}, and the explicit form of the solution is the following:
\begin{eqnarray}
ds^2& = & \left[ 1-\frac{2mr^2}{(r^2+q^2)^{3/2}}+\frac{q^2r^2}{%
(r^2+q^2)^2}\right] dt^2-\nonumber \\
& & \left[ 1-\frac{2mr^2}{%
(r^2+q^2)^{3/2}}+\frac{q^2r^2}{(r^2+q^2)^2}\right] ^{-1}dr^2-
r^2 d\Omega ^2,  
\label{regbh}
\end{eqnarray}
\begin{eqnarray}
E_r=q\,r^4\left[ \frac{r^2-5\,q^2}{(r^2+q^2)^4}+\frac{15}2\,\frac m{%
(r^2+q^2)^{7/2}}\right] .  
\label{eq:E}
\end{eqnarray}
By means of the substitution $x=r/|q|$ we can rewrite $g_{tt}$ and $E_r$ as follows 
\begin{equation}
g_{tt}=A(x,s)\equiv 1-\frac 1s\frac{x^2}{(1+x^2)^{3/2}}+\frac{x^2}{(1+x^2)^2%
},  \label{A}
\end{equation}
\beq
E_r= \frac{x^4}{q}\left[ \frac{x^2-5}{(x^2+1)^4}+\frac{15}{4s}\frac{1}{(x^2+1)^{7/2}}\right].
\label{Er}
\eeq
The result of the analysis made in \cite{prl} is that this metric describes a
 regular black
 hole. The position of the horizons was identified there with the 
values of 
the coordinate $x$  for which $g_{tt}$ is zero. These are given by
\beq
s = \frac{x^2\sqrt{x^2+1}}{x^4+3x^2+1}.
\label{zerosa}
\eeq
Accordingly, the solution 
has two horizons (for $0<s<0.317$), one horizon (for $s=
0.317$), or no horizons (for $s>0.317$). It was also stated that this 
solution is regular, on the basis of the finiteness of the three invariants 
$R$, $R_{\mu\nu}R^{\mu\nu}$, and $R_{\mu\nu\alpha\beta}R^{\mu\nu\alpha\beta}$
\footnote{ We have checked that all the components of $R_{ABCD}$ and $C_{ABCD}$ w.r.t a static observer are finite at $r=0$.}. 

Let us point out now some features of the solution described by Eqns.(\ref{A}) and (\ref{Er}) that were not noticed in \cite{prl}. First, the behaviour of the radial component of the electric field depends on the value of $s$. Specifically, $E_r$ 
may have a zero; its position is given by 
\beq
s = -\frac{15}{4}\frac{\sqrt{x^2+1}}{x^2-5}.
\label{ezero}
\eeq
Consequently, $E_r$ does not have zeros for $0<s<3/4$. 
For $s\geq 3/4$, $E_r$ has one zero located in the interval $(0, \sqrt{5})$ of the coordinate $x$. 
These features of the electric field are depicted in Figure 1 \footnote{The plots in this paper have been done with \texttt{gnuplot} \cite{gnu}.} .
\begin{figure}[h]
\centerline{\psfig{file=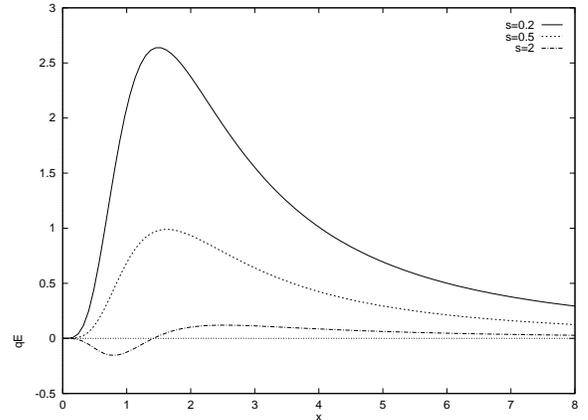,width=8cm,angle=-90}}
\caption{Electric field times the electric charge $q$ as a function of $x$ for different values of $s$.}
\label{fig1}
\end{figure}
Another salient feature of $E_r$ is that its energy density, calculated as the $G^t_t$ component of the Einstein tensor \footnote{ This and other calculations in this paper were done with the package {\em Riemann} \cite{riemann}.} may be negative for some interval of $x$. In fact, the expression  
\beq
G^t_t = \rho = \frac{1}{sq^2}\frac{s\sqrt{1+x^2}\;(x^2-3)+3(x^2+1)}{(1+x^2)^{7/2}}
\eeq
is zero for
\beq
s= -3\frac{\sqrt{1+x^2}}{x^2-3}.
\label{zeroe}
\eeq
For $s<1$, the energy is always positive, but for $s\geq 1$ it has a zero given by Eq.(\ref{zeroe}). Figure 2 illustrates the situation.
\begin{figure}[h]
\centerline{\psfig{file=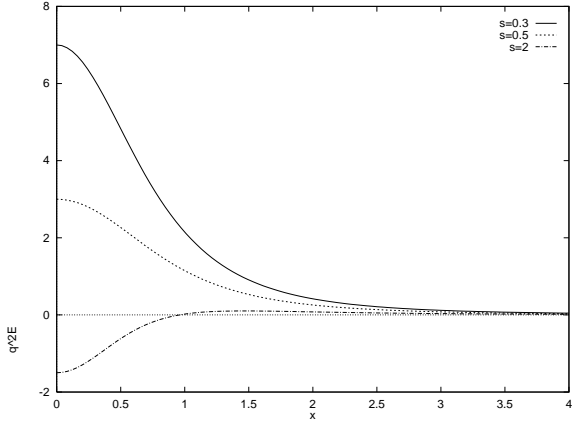,width=8cm,angle=-90}}
\caption{Energy density of the electric field times $q^2$ as a function of $x$ for different values of $s$.}
\label{fig2}
\end{figure}

\section{Effective geometry for photons}
\label{geometry}

In this section we give a summary of the method of the effective geometry \cite{nov1}. 
We will deal here  only with the case in which the Lagrangian of the nonlinear electromagnetic theory is a function of $F$ only. The general case in which ${\cal L}$ 
depends also on $G=\frac{1}{2}F^{\mu\nu}\eta_{\mu\nu}^{\alpha\beta}F_{\alpha\beta}$ is analyzed in \cite{nov1}. 
Based on the framework introduced by Hadamard \cite{had}, Novello {\em et al} showed that the discontinuities of the electromagnetic field propagate 
according to the equation
\begin{equation}
\left({\cal L}_F\eta^{\mu\nu} - 4{\cal L}_{FF}F^{\mu\alpha}F_\alpha{}^\nu\right) 
k_\mu k_\nu = 0 ,
\label{gww4}
\end{equation}
where $\eta_{\mu\nu}$ is the (flat) background metric, and $k^\mu$ is the propagation vector. This expression suggests 
that the 
self-interaction of the field $F^{\mu\nu}$ can be interpreted
as a modification on the spacetime metric 
$\eta_{\mu\nu}$, leading to the effective geometry  
\begin{equation}
g^{\mu\nu}_{\rm (eff)} = {\cal L}_{F}\,\eta^{\mu\nu}  - 
4\, {\cal L}_{FF} \,{F^{\mu}}_{\alpha} \,F^{\alpha\nu}.
\label{geffec}
\end{equation}
Note that only in 
the particular case of linear Maxwell electrodynamics 
the discontinuities of the  electromagnetic field 
propagate along the  null cones of the Minkowskian background.

The general expression of the effective geometry can be
equivalently written in terms of the energy-momentum tensor, given by 
\begin{equation}
T_{\mu\nu} \equiv \frac{2}{\sqrt{-\gamma}} 
\,\frac{\delta\,\Gamma}{\delta\,\gamma^{\mu\nu}},
\end{equation}
where $\Gamma$ is the effective action
\begin{equation}
\Gamma \doteq \int \,d^{4}x \sqrt{-\gamma}\,L,
\end{equation}
and $\gamma_{\mu\nu}$ is the Minkowski metric written in an arbitrary 
coordinate system; $\gamma$ is the corresponding determinant. 
In the case of one-parameter Lagrangians, ${\cal L}={\cal L}(F),$ we obtain
\begin{equation}
T_{\mu\nu} = - 4 {\cal L}_{F}\, {F_{\mu}}^{\alpha} \, F_{\alpha\nu} - 
{\cal L}\,\eta_{\mu\nu},  
\end{equation}
where we have chosen an Cartesian coordinate system in 
which $\gamma_{\mu\nu}$ reduces to $\eta_{\mu\nu}.$ 
In terms of this tensor the effective geometry (\ref{geffec}) 
can be re-written as%
\begin{equation}
\label{gT}
g^{\mu\nu}_{\rm (eff)}=\left({\cal L}_F+\frac{{\cal L}\,{\cal L}_{FF}}{{\cal L}_F}\right)
\eta^{\mu\nu} +\frac{{\cal L}_{FF}}{{\cal L}_F}T^{\mu\nu}.
\end{equation}
It is shown in \cite{nov1} that the field discontinuities propagate along the null geodesics
of the effective geometry given by
Eq.(\ref{gT}). This equation explicitly shows that
the stress-energy distribution of the field is the true responsible for 
the deviation of the geometry felt by photons, 
from its Minkowskian form \footnote{For $T_{\mu\nu} = 0$, the conformal 
modification in (\ref{gT}) clearly leaves the photon paths unchanged.}.
  
 We will show now that the modification of the 
underlying spacetime geometry seen by photons due to nonlinear electrodynamics 
can be also described as if
photons governed by Maxwell 
electrodynamics were propagating inside a dielectric medium. In this last case, 
the electromagnetic field is represented by two antisymmetric
tensors, the electromagnetic field $F_{\mu\nu}$ and 
the polarization field $P_{\mu\nu}$. For electrostatic fields inside 
isotropic dielectrics it follows that $P_{\mu\nu}$ and 
$F_{\mu\nu}$ are related by 
\begin{equation}
\label{rel}
P_{\mu\nu} = \epsilon(E) F_{\mu\nu}.
\end{equation}
where $\epsilon$ is the electric susceptibility. Comparing with Eq.(\ref{iden}) we see that we can make the identification
\begin{equation}
{\cal L}_{F}  \longrightarrow \epsilon,
\end{equation}
which implies 
\begin{equation}
{\cal L}_{FF} \longrightarrow -\,\frac{\epsilon\rq}{4\,E},
\end{equation}
in which $\epsilon'\equiv{d\,\epsilon}/{d\,E}$ and $E^2 \equiv -\,E_{\alpha}\,E^{\alpha} > \,0.$  
Therefore,  every Lagrangian ${\cal L} = {\cal L}(F)$ which describes a nonlinear electromagnetic theory 
may be used as a convenient description of Maxwell theory 
inside isotropic nonlinear dielectric media. 
Conversely, results obtained in the latter context 
can be restated in terms of Lagrangians of nonlinear theories.  
Using this equivalence, the effective geometry can be rewritten as
\begin{equation}
\label{gef22}
g^{\mu\nu}_{\rm (eff)} = \epsilon\, \eta^{\mu\nu} - \frac{\epsilon\rq}{E} \left( 
E^{\mu}\,E^{\nu} - E^2\, \delta^{\mu}_t \,\delta^{\nu}_t \right) .
\end{equation}
In other words,
\begin{eqnarray}
\label{gef223}
g^{tt}_{\rm (eff)} &=& \epsilon  + \epsilon' E ,\\
\label{gef224}
g^{ij}_{\rm (eff)} &=& -\,\epsilon\,\delta^{ij} - \frac{\epsilon'}{E}\, E^i\,E^j .
\end{eqnarray}
This shows that the discontinuities of the electromagnetic field 
inside a nonlinear dielectric medium propagate along null cones 
of an effective geometry (given by Eqn.(\ref{gef22})) which depends on the characteristics 
of the medium.  

Although in \cite{nov1} the background was flat, the method can also be used in a curved background. The reason is that the equations given in \cite{nov1} are valid locally in any curved spacetime. Then from the Equivalence Principle follows that the only change in Eq.(\ref{geffec}) is that of $\eta _{\mu\nu}$ by $g_{\mu\nu}$.

\section{Analysis of the ``regular black hole''}
\label{anal}

Using Eqns.(\ref{gef223}) and (\ref{gef224}) it follows that the effective metric associated to a spherically symmetric solution of Einstein's equations 
is given by
\beq
ds^2=\frac {1} {\Phi (r)} \left[ A(r) dt^2-
A(r)^{-1} dr^2\right]
-\frac{r^2}{{\cal L}_F} d\Omega ^2,  
\label{gbheff}
\end{equation}
where 
\beq
\Phi = \epsilon +\frac{d\epsilon}{dE_r} E_r = -\frac{2q}{r^3}\frac{1}{\frac{dE_r}{dr}}
\label{phidef}
\eeq 
and 
\beq
\epsilon = \frac{1}{E_r} \sqrt{\frac{-P_{\mu\nu}P^{\mu\nu}}{2}} .
\eeq
For the case dealt with in the previous section,  the function $\Phi$ takes the form
\begin{eqnarray}
\Phi (x,s) = & \nonumber \\  
& \frac{8(x^2+1)^5s}{x^6(8x^4s-104sx^2+80s+45x^2\sqrt{x^2+1}-60\sqrt{x^2+1})}.
\label{phi}
\end{eqnarray}
From Eq.(\ref{gbheff}) 
we see that the $tt$ coefficient of the effective metric is given by the quotient
$g_{tt}^{\rm (eff)} = A/\Phi$. The function $\Phi ^{-1}$ has real zeros for 
\beq
s = -\frac {15}{8}\frac{\sqrt{x^2+1}\;{(3x^2-4)}}{x^4-13x^2+10}.
\label{zerosphi}
\eeq
Taking into account that $s$ must be positive, we conclude from Eqn.(\ref{zerosphi}) 
that the function $\Phi ^{-1}$ has one zero for $s<3/4$ and two zeros for $s\geq 3/4$. 
In both cases the zeros are in the interval $(0, 3.49)$ of the coordinate $x$. 

It was shown in \cite{prl} that the metric coefficient 
$g_{tt}$ given by Eq.(\ref{A}) has two zeros for $s<0.317$, one zero for $s=0.317$, 
and no zeros for $s>0.317$.  The zeros in $g_{tt}$ were identified in \cite{prl} with 
horizons . 
We see that due to the effective metric, the geometry seen by the photons 
is more complex than the geometry seen by ordinary matter. 
Taking into account the 
zeros of $A$ and those of $\Phi ^{-1}$ we conclude that $g_{tt}^{\rm (eff)}$ has 
3 
zeros for $s<0.371$, two zeros for $s = 0.371$, one zero for $0.317 < s < 3/4$, and again
two zeros for $s\geq 3/4$.  

To determine the nature of the new zeros in the metric, it is useful to study the effective potential that is felt by the photons. The symmetries of the metric imply that there are two Killing vectors and consequently, two conserved quantities:
\beq
E_0 = g_{tt} \dot t, \;\;\;\;\;\;\;{\rm and } \;\;\;\;\;\;\;h_0 = \frac{r^2}{{\cal L}_F}\dot \phi
\eeq
(the overdot means derivative w.r.t the affine parameter). Standard calculations (see for instance \cite{wald}) using $g^{(\rm eff)}_{\mu\nu}$ show that the effective potential for photons is given by
\beq
V_{\rm eff} = (1- \Phi ^{2}) \frac{E_0^2}{2} + \frac{h_0^2}{x^2}{\cal L}_F A \Phi
\label{potef}
\eeq
The explicit form of the effective potential is too involved to be displayed here. However, we note that $V_{\rm eff}$ has poles. One of them is at $x=0$, and the others are given by the expression of the poles of $\Phi$ (see Eq.(\ref{zerosphi})), and those of ${\cal L}_F$ which are given by Eq.(\ref{ezero}).
${\cal L}_F$ has no poles for $0<s<3/4$, and one pole for $s\geq 3/4$.
Leaving aside the pole at $x=0$,  it follows that 
for $s<3/4$, the effective potential has only one pole, and for $s\geq 3/4$, it has three poles. Those that originate in the singularities of the function $\Phi$ are in agreement with the extrema of the electric field, as shown by Eq.(\ref{phi}).
We give in Figs. 3, 4 and 5 plots of $V_{\rm eff}$ for different values of the relevant parameters. 
\begin{figure}[h]
\centerline{\psfig{file=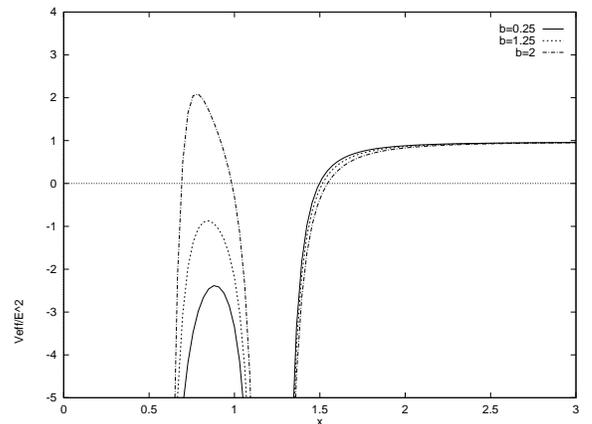,width=8cm,angle=-90}}
\caption{Effective potential $V_{\rm eff}/E_0^2$ for $s=0.2$ and different values of the impact parameter $b= h_0/E_0$. }
\label{fig3}
\end{figure}
\begin{figure}[h]
\centerline{\psfig{file=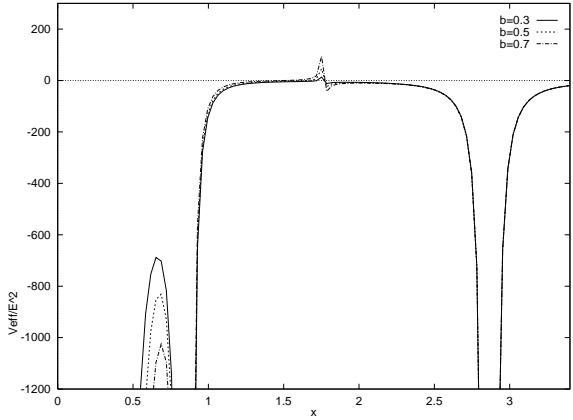,width=8cm,angle=-90}}
\caption{Effective potential $V_{\rm eff}/E_0^2$ for $s=4$ and different values of $b $. The interval $1.7\leq x \leq 1.8$ is shown in detail in the next figure.}
\label{fig4}
\end{figure}
\begin{figure}[h]
\centerline{\psfig{file=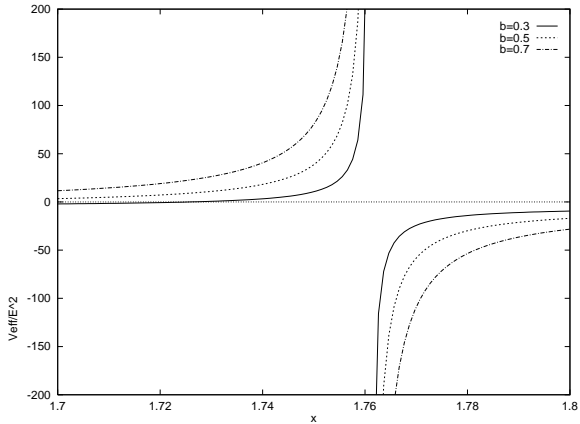,width=8cm,angle=-90}}
\caption{Effective potential $V_{\rm eff}/E_0^2$ for $s=4$ and different values of $b $. The singularity seen in this plot comes from the pole of ${\cal L}_F$.}
\label{fig4}
\end{figure}

Several comments are in order.  
The singularities in the potential suggest that the 
effective geometry itself may be singular. This is confirmed by the expression of the 
scalar curvature $R^{\rm (eff)}$, which diverges in the values of $x$ given by 
Eqs.(\ref{ezero}) and (\ref{zerosphi}). 
Let us analize the relative position of these 
singularities felt by the photons and those of the metric coefficient $g_{tt}(x,s)$, 
given by Eq.(\ref{zerosa}).
The information is conveniently summarized by the 
following plot:
\begin{figure}[h]
\centerline{\psfig{file=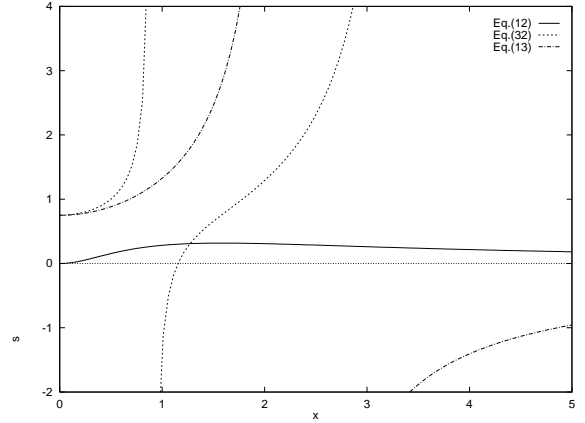,width=8cm,angle=-90}}
\caption{Position of the zeros of the metric and spacetime singularities  for a given $s$.}
\label{zeros}
\end{figure}
We see that for a fixed $s\leq 0.317 $ the singularities are situated inside the first horizon.
However, 
for $s>0.317$ the singularities are not anymore hidden behind a horizon: we are then in the presence of naked singularities. We must remark that these singularities are only felt by photons. The rest of the matter follows geodesics of the regular spacetime given in \cite{prl}.

It can also be seen from the plot that for $s<0.371$ the coordinate distance between the two horizons decreases for increasing $s$, up to $s=0.371$, where the two horizons coalesce.

Before analyzing the path of a photon coming from infinity, let us remark that there is a low potential barrier extending to the right of the outermost singularity  for any value of the parameters. This barrier can be seen in 
Fig.\ref{fig3}, and it is also present to the right of Fig.4. 
A low-energy photon incident from the right will find then this barrier, and will be deflected back to large values of $x$. This deflection will be more pronounced with increasing energy. When the energy of the photon is aproximately that of the height of the barrier, the photon can orbit around the center of the field in an unstable orbit. Finally an incident photon with energy greater that the height of the barrier will inevitably encounter the first singularity.

It is easily seen from Eq.(\ref{potef})
that the potential goes to zero for large values of $x$.
We have also analyzed the effective potential for the case of a negative $q$, but the only quantitatively different result is a small increment of the 
innnermost local maximum seen in Figs.3 and 4.

We move now to another peculiar feature of the effective geometry. It is known that the effective potential for the Schwzarschild and Reissner-Nordstrom  geometries is null in the case of photons with $h_0 = 0$. However, from Eq.(\ref{potef}) we see that in this case $V_{\rm eff}$ for the effective geometry reduces to
\beq
V_{\rm eff} = ( 1 - \Phi ^2) E_0^2
\label{potef0}
\eeq
The dependence of this potential on $\Phi$ is the same as in Eq.(\ref{potef}), so the behaviour of $V_{\rm eff}$ with $x$ in this case is qualitatively depicted in Figs.3 and 4.

Let us finally point out some unusual geometrical properties of the metric seen by the photons. 
The effective metric has the same symmetries of the original metric given by Eq.(\ref{regbh}). It can be easily shown that the time Killing vector $\partial /\partial t$ is null on the hypersurfaces determined by the zeros of $g_{tt}^{\rm (eff)}$. 

Another interesting property of these surfaces is associated to the redshift of the photons. The redshift $z$ of a source as measured by an observer with velocity $u^\mu$ can be defined in terms of the frequency by
\beq
1+z = \frac{(u^\mu k_\mu )_{\rm emitter}}{(u^\mu k_\mu)_{\rm observer}}.
\eeq
Considering a static observer for which $u^\mu = \delta ^\mu _0 / \sqrt{g_{tt}}$
this expression can be written as 
\beq
1+z = \left[ \frac{\sqrt{g_{tt}}}{g^{\rm (eff)}_{tt}}\right]_{\rm em}
\left[\frac{g_{tt}^{\rm (eff)}}{\sqrt{g_{tt}}}\right]_{\rm obs}
\eeq
Using the expression of the effective metric, and if the observer is at infinity, 
\beq
1+z = \frac{\Phi}{\sqrt{A}}
\eeq
We conclude then that the redshift diverges in two cases: when $A$ is zero, and when $\Phi$ diverges
(see Fig.(\ref{zeros})).

\section{Conclusion}

The remarkable fact that in nonlinear electrodynamics the trajectories of photons are modified by the nonlinearities of the field equations has not been addressed frequently in the literature. The photons do not propagate following the null cones of the background metric but those
of the effective metric.
We have shown here the dramatic consequences that this has in a so-called regular black hole. In this case, there are singularities that are seen only by the photons. These singularities can either be hidden behind a horizon or naked, according to the value of the ratio $q/2m$.  Let us remark that the existence of singularities in these type of solutions is a direct consequence of the existence of extrema of the electric field, as Eq.(\ref{phidef}) shows.  This is a general property which will always be present in any static and spherically symmetric solution of the system of equations (\ref{Ein}) and (\ref{max}) when the electromagnetic theory is nonlinear.

We have also shown that the effective potential to the right of the outermost singularity resembles that of Schwarzschild and Reissner-Nordstrom. However, contrary to what happens in Maxwell theory, photons with zero angular momentum travel under the influence of an effective potential that is different from zero. 

We also exhibited some unusual properties of the solution found in \cite{prl}. The electric field may have one or two extrema depending on the value of $s$. In the second case, it has a zero. Also, for certain values of $s$ the energy of the electric field is negative in some coordinate range. There are at least two more properties, geometrical in origin, that are worth of notice. First, the time Killing vector of the effective geometry is null in the surfaces where the function $\Phi$ diverges. Second, the redshift measured by an observer far from the source  
diverges on the same surfaces. It is important to remark that these geometrical  properties will be present in every solution with the same symmetries if the electric field has extrema.

To close, we would like to emphasize that ordinary matter follows geodesics of the background metric. However, the modifications of the metric induced by the nonlinearities of the electromagnetic field must always be taken into account when studying the propagation of photons. The abovementioned properties are nothing but a consequence of the nonlinearities of the electromagnetic theory.  

\acknowledgements
SEPB would like to acknowledge financial support from CONICET-Argentina. MN and JMS acknowledge financial support from CNPq. The authors would like to thank K. Bronnikov for a useful remark.

\end{document}